\documentclass[pra,twocolumn,showpacs,showkeys,preprintnumbers,amsmath,amssymb]{revtex4}

\usepackage{graphicx}
\usepackage{dcolumn}
\usepackage{bm,amsmath,amsfonts,amssymb,mathrsfs}

\newcommand{\ot}[0]{\otimes}

\newcommand{\ket}[1]{|#1\rangle}
\newcommand{\bra}[1]{\langle#1|}
\newcommand{\kett}[1]{\ket{#1}}
\newcommand{\braa}[1]{\bra{#1}}

\newcommand{\proj}[1]{\ket{#1}\bra{#1}}
\newcommand{\projj}[1]{\kett{#1}\braa{#1}}

\newcommand{\bb}[1]{\mathbb{#1}}
\newcommand{\be}[0]{\begin{equation}}
\newcommand{\ee}[0]{\end{equation}}
\newcommand{\bea}[0]{\begin{eqnarray}}
\newcommand{\eea}[0]{\end{eqnarray}}
\newcommand{\te}[1]{\tilde{e}_{#1}}

\newcommand{\tf}[1]{\tilde{f}_{#1}}
\newcommand{\pv}[1]{\ket{\te{#1}}\ot\ket{\tf{#1}}}
\newcommand{\pf}[1]{\bra{\te{#1}}\ot\bra{\tf{#1}}}

\newcommand{\Tr}[0]{\mathrm{Tr}}
\newcommand{\pbit}{\gamma_{ABA'B'}^{(2)}}

\begin{document}
\title{Quantum states representing perfectly secure bits are always distillable}

\author{Pawe\l{} Horodecki}
\email{pawel@mif.pg.gda.pl}
\author{Remigiusz Augusiak}
\email{remik@mif.pg.gda.pl}
 \affiliation{Faculty of Applied
Physics and Mathematics, Gda\'nsk University of Technology,
Gda\'nsk, Poland}

\date{\today}

\begin{abstract}
It is proven that recently introduced states with perfectly secure
bits of cryptographic key (private states representing secure bit)
[K. Horodecki {\it et al.}, Phys. Rev. Lett. {\bf 94}, 160502
(2005)] as well as its multipartite and higher dimension
generalizations always represent distillable entanglement. The
corresponding lower bounds on distillable entanglement are
provided. We also present a simple alternative proof that for any
bipartite quantum state entanglement cost is an upper bound on
distillable cryptographic key in bipartite scenario.
\end{abstract}

\maketitle
\section{Introduction}
For a long time quantum cryptography with entanglement discovered
by Ekert \cite{Ekert91} has been developed based on pure quantum
entanglement as a central resource. More precisely the schemes
(see Ref. \cite{schemes}) existing in this domain were equivalent
to entanglement distillation \cite{puryfikacja}. Recently it has
been shown that entanglement which is not distillable (bound
entanglement) can provide a quantum cryptographic key \cite{KH1}.
This leads to a general scheme of key distillation from quantum
states \cite{KH1,KH2} with a private state representing a secure
bit (alternatively: a private bit state or private bit) as an
important notion. The latter is a quantum state shared by Alice
and Bob that contains at least one bit of perfectly secure
cryptographic key. Quite nonintuitively, private bits can be
approximated arbitrary well by some bound entangled states in some
special sense: there exist a sequence of private bits with
dimension of their "shield" part going to infinity and another
sequence of bound entangled states such that trace distance
between elements of the two sequences goes to zero \cite{KH1,KH2}.
Here we show that despite of that fact any single private bit is
distillable. Using local filtering \cite{filter} in a way
exploited in Ref. \cite{pur} we provide a lower bound on the
corresponding distillable entanglement of $d$--dimensional private
state (d-dimensional generalization of p-bit). Note that the
special (bipartite) case of our result has already found an
important application in a proof of unconditional cryptographic
security with small distillable entanglement \cite{small}. Finally
we give a simple alternative proof that for any bipartite state
entanglement cost $E_{C}$ (see Refs.
\cite{EntanglementCost,Michal}) is an upper bound on amount of a
distillable cryptographic key $K_{D}$ (usually called distillable
key). Originally this was proven \cite{KH1,KH2} from the fact that
regularized entropy of entanglement $E_{R}^{\infty}$ is an upper
bound for $K_{D}$. We provide a simpler version that does not need
to refer to $E_{R}^{\infty}$.

%
\section{Lower bounds on distillability}
\subsection{Distillation of entanglement from private bits}
For purely pedagogical reasons at the very beginning we shall
derive the lower bound on distillable entanglement of private
bits. This is the starting point for more general classes of
secure states as $d$--dimensional private states and their
multipartite counterparts.

Let us recall the definition of private bit \cite{KH1,KH2}. This
is the bipartite state with internal structure of both Alice and
Bob systems. It is defined on four-partite Hilbert space
$\mathcal{H}_{ABA'B'}=\mathcal{H}_{A} \ot \mathcal{H}_{B}\ot
\mathcal{H}_{A'} \ot \mathcal{H}_{B'}$ with subsystems denoted by
$A$ and $A'$ ($B$ and $B'$) belonging to Alice (Bob). The first
pair of subsystems shared by Alice and Bob is of qubit structure,
i.e., $\mathcal{H}_{AB}=\mathcal{H}_{A}\ot
\mathcal{H}_{B}\sim\bb{C}^{2}\ot\bb{C}^{2}$, while the second one
has in general the form  $\mathcal{H}_{A'B'}=\mathcal{H}_{A'}\ot
\mathcal{H}_{B'}\sim\bb{C}^{d_{A'}}\ot\bb{C}^{d_{B'}}$. The
explicit form of private bit is
%
\begin{equation}
\pbit\equiv\gamma^{(2)}=\frac{1}{2}\sum_{i,j=0}^{1}\ket{ii}\bra{jj}\ot
U_{i}\rho_{A'B'}U_{j}^{\dagger}, \label{pbit}
\end{equation}
where $\{ |ij\rangle\} $ is the standard two-qubit product basis
in ${\cal H}_{AB}$, $\rho_{A'B'}$ denotes some state acting on
$\mathcal{H}_{A'B'}$ and $U_{i}$ $(i=0,1)$ are some unitary
operations. The structure of private bit can be divided into two
parts \cite{KH2}. The first one $(AB)$ called {\it key part} is
the one from which Alice and Bob can get a bit of secure key after
local measurements in the standard bases. The second part $(A'B')$
is called {\it shield part} (in the case of approximate private
bits this part in a sense "defends" the key in system $AB$ from an
eavesdropper in the asymptotic regime \cite{KH2}). We must stress
that the private bit contains the {\it perfectly} secure bit of
key while it can be approximated by bound entangled states in the
sense that there exist a sequence of private bits
$\gamma^{(2)}_{k}$ with the dimension of their shield parts
$\mathcal{H}_{A'B'}^{(k)}$ going to infinity and another sequence
of bound entangled states $\varrho_{k}$ such that for any
$\epsilon$ there exist $k$ such that
$||\gamma^{(2)}_{k}-\varrho_{k}||\leq \epsilon$. The latter has
been proven to imply that a bound entangled state can contain {\it
secure bit up to arbitrary precision} measured by $\epsilon$ in
the sense that the eavesdropper information about the shared bit
is bounded by some continuous function of $\epsilon$ that vanishes
for $\epsilon=0$ (see Ref. \cite{KH2}). In general one has a
perfectly secure bit (no Eve's knowledge about the bit) for
$\epsilon=0$, i.e., when the observers share just a private bit.
It turns out that this exact case can never happen when the
observers are given bound entanglement since, according to main
result of the present paper, a private bit is always distillable.

Having reviewed the structure of $\gamma^{(2)}$, we can start a
description of distillation protocol. We define the following
parameter
\begin{equation}\label{eta}
\eta=\max\left|\bra{e_{1}}\ot\bra{f_{1}}U_{0}\rho_{A'B'}
U_{1}^{\dagger}\ket{e_{2}}\ot\ket{f_{2}}\right|,
\end{equation}
where the maximum is taken over all normalized product vectors
$\ket{e_{1}}\ot\ket{f_{1}}$ and $\ket{e_{2}}\ot \ket{f_{2}}$
belonging to $\mathcal{H}_{A'B'}$.

One immediately infers that
\begin{equation*}\label{PositivityEta}
\eta \ge
\max_{\substack{m,n=0,\ldots,d_{A'}-1\\\mu,\nu=0,\ldots,d_{B'}-1}}
\left|\left[U_{0}\rho_{A'B'}U_{1}^{\dagger}\right]_{m\mu,n\nu}\right|>0,
\end{equation*}
where strict positivity follows from the fact that the matrix
representation $[U_{0}\rho_{A'B'}U_{1}^{\dagger}]_{m\mu,n\nu}$ of
the nonzero operator $U_{0}\rho_{A'B'}U_{1}^{\dagger}$ in standard
basis $\{| ij \rangle \}$ must have at least one nonzero element.
Let $\pv{1}$ and $\pv{2}$ be product vectors from
$\mathcal{H}_{A'B'}$ for which the maximum in (\ref{eta}) is
achieved. Then we define numbers
\begin{equation*}\label{as}
a_{s}=\pf{s}U_{s-1}\rho_{A'B'}U_{s-1}^{\dagger}\pv{s}
\qquad(s=1,2).
\end{equation*}
These numbers are always positive and the square root of their
product is bounded from below by $\eta$ (see Ref. \cite{Append}).
Now we are in a position to show the distillability of
$\gamma^{(2)}$. Let us assume that $a_{2}\ge a_{1}>0$. Then we
define local operators:
\begin{equation*}
V_{AA'}=\proj{0}\ot\bra{\te{1}}+\sqrt{\frac{a_{1}}{a_{2}}}\,\mathrm{e}^{\mathrm{i}\Theta}\proj{1}\ot\bra{\te{2}},
\end{equation*}
with
\begin{equation*}
\Theta=\arg\left[\bra{\tilde{e}_{1}}\ot\bra{\tilde{f}_{1}}U_{0}\rho_{A'B'}U_{1}^{\dagger}
\ket{\te{2}}\ot\ket{\tf{2}}\right],
\end{equation*}
and
\begin{equation*}
P_{BB'}=\proj{0}\ot\bra{\tf{1}}+\proj{1}\ot\bra{\tf{2}}.
\end{equation*}
The above operators can be used in LOCC operation of two-way type.
The operation (called two-way local filtering) produces with
probability $a_{1}>0$ the state:
\begin{eqnarray}\label{mixture}
\varrho&\equiv&\frac{V_{AA'}\ot P_{BB'}\pbit V_{AA'}^{\dagger}\ot
P_{BB'}^{\dagger}}{\Tr\left[V_{AA'}\ot P_{BB'}\pbit
V_{AA'}^{\dagger}\ot
P_{BB'}^{\dagger}\right]}\nonumber\\
&=& p\proj{\Psi_{+}}+(1-p)\proj{\Psi_{-}},
\end{eqnarray}
with $p=(1/2)(1+\eta/\sqrt{a_{1}a_{2}})$ and two Bell states
$\ket{\Psi_{\pm}}=(1/\sqrt{2})(\ket{00}\pm\ket{11})$. Distillable
entanglement of two-element mixture of Bell states is known to be
\cite{huge,Rains} $E_{D}(\varrho)=1-H(p)$, where $H(p)=-p\log
p-(1-p)\log(1-p)$ and can be achieved in the so--called hashing
protocol \cite{huge}. If $a_{1}\ge a_{2}$ one applies the same
procedure with only one modification, i.e., putting the local
filter
$W_{AA'}=\sqrt{a_{2}/a_{1}}\mathrm{e}^{-i\Theta} V_{AA'}$ in place
of $V_{AA'}$. The resulting state is equal to the same mixture of
Bell states (\ref{mixture}) as in the previous case, but the
probability of its production is now $a_{2}$. Combining these two
observations we have the lower bound on distillable entanglement
of $\gamma^{(2)}$:
\begin{equation}\label{lboundbit}
E_{D}\big(\gamma^{(2)}\big)\ge
a_{\max}\left[1-H\left(\frac{1}{2}+\frac{\eta}{2\sqrt{a_{1}a_{2}}}\right)\right],
\end{equation}
where the factor $a_{\max}=\max[a_{1},a_{2}]$ is maximum of  two
probabilities of production of the considered two-qubit Bell
states mixture. It should be emphasized that $\eta > 0$ and
therefore the Shannon entropy in (\ref{lboundbit}) is less than
one, which results in strict positivity of right--hand side of
(\ref{lboundbit}). Thus a bipartite private bit is always a
distillable state.
 \\
%

\subsection{Distillability of multipartite p-dits}
%
Here we shall provide a generalization of the result to any
multipartite version of a bipartite $d$--dimensional private state
(hereafter denotes by $\gamma_{ABA'B'}^{(d)}$ or shortly by
$\gamma^{(d)}$) \cite{KH1,KH2}. Multipartite $d$--dimensional
private states play a natural role in the generalized scheme of
distillation of a secure key in a multipartite scenario
\cite{MultiCryptoBE}. As mentioned, the special case of the
present result, i.e., the $\gamma^{(d)}$ one has already been
applied in an unconditional security proof with a small
distillable entanglement \cite{small}.

A multipartite $d$--dimensional private state is a natural
generalization of the private bit (\ref{pbit}) both in the
''size'' of the key part (increased for any local observer from
dimension $2$ to $d$; this leads to $\log d$ of secure bits of key
\cite{KH1}) and in the number of observers involved
\cite{MultiCryptoBE}: from two observers Alice ($AA'$) and Bob
($BB'$) to $N$ ones $\{
(A_{1}A_{1}'),(A_{2}A_{2}'),\ldots,(A_{N}A_{N}') \}$. It obviously
reproduces the bipartite $d$--dimensional private state \cite{KH1}
in case of two observers. The form of multipartite ($N$-partite)
$d$--dimensional private state is
\begin{equation*}\label{mpdits}
\Gamma_{\textsf{A}\textsf{A}'}^{(d)}=\frac{1}{d}\sum_{i,j=0}^{d-1}\ket{i\ldots
i}\bra{j\ldots j}\ot U_{i}\varrho_{\textsf{A}'}U_{j}^{\dagger},
\end{equation*}
The above state is defined on a Hilbert space
$\mathcal{H}_{\textsf{A}\textsf{A}'}=\mathcal{H}_{\textsf{A}}\otimes
\mathcal{H}_{\textsf{A}'} \equiv
(\mathcal{H}_{A_{1}}\ot\ldots\ot\mathcal{H}_{A_{N}}) \otimes
(\mathcal{H}_{A_{1}'}\ot\ldots\ot\mathcal{H}_{A_{N}'})$. Here the
system  $\textsf{A}=A_{1}\ldots A_{N}$ is of $d ^{\otimes N}$ type
(one has $N$ systems of $d$-level type instead of two systems of
qubit type) with the standard basis $\{ |i_{1} \ldots i_{N}
\rangle\}$. The density matrix $\varrho_{A_{1}'\ldots A_{N}'}$
acting on a Hilbert space $\mathcal{H}_{\textsf{A}'}$ is
responsible for the shield part of $\Gamma_{\mathsf{AA}'}^{(d)}$
and, as previously, $U_{i}$ $(i=0,\ldots,d-1)$ are certain unitary
evolutions.

The distillation scheme may be found using similar techniques as
for private bits. Therefore for fixed $i$ and $j$ $(i<j,\;
i,j=0,\ldots,d-1)$ let us define
\begin{equation}\label{mpditEta}
\eta^{(ij)}=\max\left|\braa{f_{1}}\ot\ldots\ot\braa{f_{N}}
U_{i}\varrho_{\textsf{A}'}U_{j}^{\dagger}
\kett{g_{1}}\ot\ldots\ot\kett{g_{N}}\right|,
\end{equation}
where maximum is taken over all product vectors from
$\mathcal{H}_{\textsf{A}'}$. Similarly as in the private bit case
we also define
\begin{equation*}\label{ampdits1}
a_{1}^{(ij)}=\bra{\tilde{f}_{1}^{(ij)}}\ot\ldots\ot\bra{\tilde{f}_{N}^{(ij)}}
U_{i}\varrho_{\textsf{A}'}U_{i}^{\dagger}\ket{\tilde{f}_{1}^{(ij)}}\ot\ldots\ot
\ket{\tilde{f}_{N}^{(ij)}},
\end{equation*}
and
\begin{equation*}\label{ampdits2}
a_{2}^{(ij)}=\bra{\tilde{g}_{1}^{(ij)}}\ot\ldots\ot\bra{\tilde{g}_{N}^{(ij)}}
U_{j}\varrho_{\textsf{A}'}U_{j}^{\dagger}\ket{\tilde{g}_{1}^{(ij)}}\ot\ldots\ot\ket{\tilde{g}_{N}^{(ij)}}
\end{equation*}
where $ \ket{\tilde{f}_{1}^{(ij)}}\ot\ldots\ot
\ket{\tilde{f}_{N}^{(ij)}}$ and $
\ket{\tilde{g}_{1}^{(ij)}}\ot\ldots\ot\ket{\tilde{g}_{N}^{(ij)}}$
are vectors realizing a maximum in Eq. (\ref{mpditEta}). In a full
analogy to the case of a private bit, one checks that (cf.
\cite{Append}) $0<\eta^{(ij)}\le \sqrt{a^{(ij)}_{1}a^{(ij)}_{2}}$.
Again, if for a given pair of indices $\{i,j\}$ ($i<j$) one has
%
$a^{(ij)}_{2}\ge a^{(ij)}_{1}>0$, we define
\begin{equation*}
V_{A_{1}A_{1}'}^{(ij)}=\proj{i}\ot\bra{\tilde{f}_{1}^{(ij)}}+\sqrt{a_{1}^{(ij)}/a_{2}^{(ij)}}
\mathrm{e}^{\mathrm{i}\Theta_{ij}}\proj{j}\ot\bra{\tilde{g}_{1}^{(ij)}},
\end{equation*}
where
\begin{equation*}
\Theta_{ij}=\arg\left[\bra{\tilde{f}_{1}^{(ij)}}\ldots\bra{\tilde{f}_{N}^{(ij)}}
U_{i}\varrho_{\mathsf{A}'}U_{j}^{\dagger}\ket{\tilde{g}_{1}^{(ij)}}\ldots\ket{\tilde{g}_{N}^{(ij)}}\right],
\end{equation*}
while in the case $a^{(ij)}_{1}\ge a^{(ij)}_{2}> 0$ we take
$W_{A_{1}A_{1}'}^{(ij)}=\sqrt{a^{(ij)}_{2}/a^{(ij)}_{1}}\,
\mathrm{e}^{-i\Theta_{ij}} V_{A_{1}A_{1}'}^{(ij)}$.
Finally we introduce
\begin{equation*}
P_{A_{k}A_{k}'}^{(ij)}=\proj{i}\ot\bra{\tilde{f}_{k}^{(ij)}}
+\proj{j}\ot\bra{\tilde{g}_{k}^{(ij)}}\quad(k=2,\ldots,N).
\end{equation*}
In both cases for given $i$ and $j$ $(i<j)$, we shall obtain the
same state but with different probabilities, $a^{(ij)}_{1}$ in the
first case and $a^{(ij)}_{2}$ in the second one. The corresponding
LOCC filtering performed by all $N$ parties (the first party uses
$V_{A_{1}A_{1}'}^{(ij)}$ or $W_{A_{1}A_{1}'}^{(ij)}$ while all the
others apply $P_{A_{k}A_{k}'}^{(ij)}$ in full analogy to the
formula (\ref{mixture})) finally gives the state
\begin{equation}
\label{mix2}
\hspace{-0.01cm}\varrho_{N}^{(ij)}=p^{(ij)}\proj{\tilde{\Psi}^{(ij)}_{+}}
+(1-p^{(ij)})\proj{\tilde{\Psi}^{(ij)}_{-}},
\end{equation}
which is the mixture of two projectors onto GHZ states
$\ket{\tilde{\Psi}^{(ij)}_{\pm}}=(1/\sqrt{2})(\ket{i\ldots
i}\pm\ket{j\ldots j})$ with
$p^{(ij)}=(1/2)\big[1+\eta^{(ij)}/(a_{ij}^{(1)}a_{ij}^{(2)})^{1/2}\big]$.
Using the GHZ distillation hashing protocol \cite{Maneva} (which
is a generalization of that from \cite{huge}) and taking into
account the fact that here one has only the so--called phase error
(corresponding to the sign $\pm$ in the above formula) we get
lower bound for the distillation rate of the GHZ states from
$\Gamma^{(d)}_{\mathsf{AA'}}$ in the scenario with chosen
filtering corresponding to a fixed pair of indices $\{ i,j \}$ as
below
\be E_{D}^{(ij)}\big(\Gamma^{(d)}_{\mathsf{AA'}}\big)\ge
\label{rate1} a_{\max}^{(ij)}\left[1-H\left(\frac{1}{2}
+\frac{\eta^{(ij)}}{\sqrt{a^{(ij)}_{1}a^{(ij)}_{2}}}\right)\right]
\equiv \tilde{E}_{D}^{(ij)}\nonumber \ee
with $a^{(ij)}_{\max}$ being the bigger from two numbers
$a^{(ij)}_{1}$ and $a^{(ij)}_{2}$. Again, since all $\eta^{(ij)}$
are positive the above lower bounds are strictly positive too,
which results in distillability of GHZ from {\it any} multipartite
$d$--dimensional private state. Since we can optimize over choices
of $\{ i,j \}$ we get the final lower bound on distillable
entanglement of $\Gamma_{\mathsf{AA}'}^{(d)}$
\begin{equation*} \label{rate2} E_{D}\big(\Gamma_{\mathsf{AA'}}^{(d)}\big)
\ge\max_{\substack{i,j=0,\ldots,d-1\\(i<j)}}\tilde{E}_{D}^{(ij)},
\end{equation*}
which again is strictly positive since, as previously proven, all
quantities $\tilde{E}_{D}^{(ij)}$ are strictly positive.

The above protocols are working for any dimensions $d$. However,
for $d\ge 4$, further generalization of efficiency of distillation
protocol can be introduced. This is because all local projections
are here two-dimensional. It is easy to generalize the above
scheme in such a way that instead of single filtering of that type
we perform POVM involving $k$ filterings ($2k\geq d$) which are
locally orthogonal in the sense that their supports (subspaces on
which they give nonzero results) are {\it disjoint}. Each of the
results corresponding to $k$th result of POVM would produce some
mixture of type (\ref{mix2}). Such a scheme would be, to some
extent, analogous to distillation of entanglement from mixtures of
locally orthogonal states \cite{APS} which was independently
analyzed also in \cite{Werner}.

%
\section{Bound on distillable key: an alternative proof}
%
In this section we come back again to the bipartite scenario. We
bound from above the amount of a distillable cryptographic key
$K_{D}(\varrho)$ of any given state $\varrho$ by its entanglement
cost $E_{C}(\varrho)$. This fact has already been proven in Ref.
\cite{KH2} through the regularized relative entropy of
entanglement. The present proof has a more direct character and is
based on the well--known facts from the theory of entanglement
measures \cite{Michal}. It also exploits the special structure of
the eigenvectors of the bipartite $d$--dimensional private states.

The crucial role is played here by the asymptotic continuity of
entanglement of formation \cite{huge} proved by Nielsen
\cite{Nielsen} and the fact that the entanglement cost which has a
rather complicated definition (see \cite{Michal}) may be related
to entanglement of formation in a simple way by
\cite{EntanglementCost}
\begin{equation}\label{DefEC}
E_{C}(\rho)=\lim_{m\to\infty}\frac{E_{F}(\rho^{\ot m})}{m}
\end{equation}
for any given state $\rho$. Moreover, for these two measures we
have $E(\Lambda(\rho))\le E(\rho)$ with $\Lambda$ being some LOCC
protocol \cite{LOCC}.

At the very beginning we show that entanglement of formation of
$\gamma_{ABA'B'}^{(d)}$ \cite{KH1} corresponding here just to
bipartite version of $\Gamma_{\mathsf{AA'}}^{(d)}$ with $d$ being
the dimension of $\mathcal{H}_{A}$ (equivalently
$\mathcal{H}_{B}$), may be bounded from below by $\log d$. This
can be obtained simply by utilizing the definition of entanglement
of formation which for a given density matrix $\rho$ acting on the
Hilbert space $\mathcal{H}_{A}\ot\mathcal{H}_{B}$ reads
\begin{equation*}
E_{F}(\rho)=\min_{\{p_{i},\kett{\Psi_{i}}\}}\sum_{i}p_{i}S_{\rm
vN}\left(\Tr_B\proj{\Psi_{i}}\right),
\end{equation*}
where the minimum is taken over all ensembles
$\{p_{i},\kett{\Psi_{i}}\}$ generating the state $\rho$ and
$S_{\rm vN}$ stands for the von Neumann entropy \cite{vNEntropy}.

Consider now the state $\gamma^{(d)}_{ABA'B'}$. One may easily see
that all its eigenvectors corresponding to nonzero eigenvalues are
\begin{eqnarray*}
\kett{\psi_{k}}&=&\frac{1}{\sqrt{d}}\sum_{j=0}^{d-1}\kett{jj}\ot
U_{j}\kett{\varphi_{k}^{(A'B')}}\nonumber\\
&=&\frac{1}{\sqrt{d}}\sum_{j=0}^{d-1}\kett{jj}\ot
\kett{\varphi_{j,k}^{(A'B')}}.
\end{eqnarray*}
An arbitrary vector $\kett{\Psi_{i}}$ from any ensemble
$\{p_{i},\kett{\Psi_{i}}\}$ realizing the considered state must be
a linear combination of the above eigenvectors $\kett{\psi_{k}}$.
This is a consequence of the fact, proved in Ref. \cite{PH}, that
the vector $\kett{\Psi_{i}}$ belongs to
$\mathrm{Ran}\gamma^{(d)}$, which in turn is a subspace spanned by
eigenvectors $\kett{\psi_{k}}$. Therefore it is not difficult to
see that $\kett{\Psi_{i}}=(1/\sqrt{d})\sum_{j=0}^{d-1}\kett{jj}
\ot \kett{\tilde{\varphi}_{j,k}^{(A'B')}}$ must hold for some
vectors $\kett{\tilde{\varphi}_{j,k}^{(A'B')}}$. Entanglement of
formation of the vector $\kett{\Psi_{i}}$ can be easily estimated
\begin{eqnarray*}
E_{F}(\kett{\Psi_{i}})&=&S_{\rm vN}\big(\Tr_{BB'} \projj{\Psi_{i}}\big)= \nonumber \\
&=&\frac{1}{d} \sum_{j} S_{\rm vN} \big(\Xi^{(A')}_{j,k}\big)+\log
d
\nonumber\\
&\ge& \log d,
\end{eqnarray*}
where
$\Xi^{(A')}_{j,k}=\Tr_{B'}\projj{\tilde{\varphi}_{j,k}^{(A'B')}}$.
Since this holds for any vector from any ensemble of
$\gamma^{(d)}$, we have, by the very definition of $E_{F}$, the
following

{\it Property 1. For any p-dit state $\gamma^{(d)}$ one has}
\begin{equation}\label{FormBound}
E_{F}\big(\gamma^{(d)}\big)\ge \log d.
\end{equation}

Note by the way that, by inspection, one can see that tensor
product $(\gamma^{(d)})^{\otimes m}$ has a structure of
$\gamma^{(d^{m})}$ type. Hence, a straightforward application of
(\ref{DefEC}) to $\gamma^{(d)}$ in place of $\varrho$ together
with the above inequality leads to a stronger result, namely one
has

 {\it Property 1a. For any p-dit state
$\gamma^{(d)}$ one has:}
\begin{equation*}\label{FormBound0}
E_{C}\big(\gamma^{(d)}\big)\ge \log d.
\end{equation*}

Now we are in position to relate $K_{D}$ and $E_{C}$. Suppose that
Alice and Bob share $n$ copies of a given bipartite state
$\varrho$ with $K_{D}(\varrho)>0$ (for states with
$K_{D}(\varrho)=0$ the inequality is trivially true). Consider the
optimal protocol distilling $K_{D}(\varrho)$ secret bits from
$\varrho$ which is a sequence of LOCC protocols $\Lambda_{n}$
($n\in\mathbb{N}$) such that $\Lambda_{n}\big(\varrho^{\ot
n}\big)=\sigma^{(n)}$ and
$||\sigma^{(n)}-\gamma^{(d_{n})}||_{\Tr}\le \epsilon_{n}$ for
sequence of private states $\gamma^{(d_{n})}$ with the key part
$AB$ defined on the Hilbert space $\mathbb{C}^{d_{n}} \otimes
\mathbb{C}^{d_{n}}$. The sequence $\{\epsilon_{n}\}$ is supposed
to converge to zero with increasing $n$. Since the protocol is
optimal we have by definition \cite{KH2} $K_{D}(\varrho)=\lim_{n}
(\log d_n/n)$.

On the other hand the asymptotic continuity of $E_{F}$ together
with its monotonicity under LOCC protocol implies the following
inequalities:
\begin{eqnarray}\label{ascontForm}
\hspace{-0.5cm}\frac{1}{n}E_{F}\big(\gamma^{(d_{n})}\big)&\le&\frac{1}{n}E_{F}\big(\sigma^{(n)}\big)+c\epsilon_{n}\log
d_{n} + \frac{o(\epsilon_{n})}{n}\nonumber\\
\hspace{-0.5cm}&\le&\frac{1}{n}E_{F}\big(\varrho^{\ot n}\big)+c
\epsilon_{n} \log d_{n} + \frac{o(\epsilon_{n})}{n}\nonumber
\end{eqnarray}
for some constant $c$ and $o(\epsilon_{n})$ vanishing faster than
$\epsilon_{n}$ in the limit of large $n$.
Combining this with (\ref{FormBound}) one has
\begin{eqnarray*}\label{ascontForm2}
\frac{\log d_{n}}{n} &\le&\frac{1}{n} E_{F}\big(\varrho^{\ot
n}\big)+c\epsilon_{n} \frac{\log d_{n}}{n} +
\frac{o(\epsilon_{n})}{n}.
\end{eqnarray*}
Taking the limit on both sides, utilizing Eq. (\ref{DefEC}), and
exploiting the fact that $K_{D}=\lim_{n} (\log d_{n}/n)$, we get
finally the desired inequality that can be stated as follows (see
\cite{KH1,KH2} for alternative proof):
%

{\it Property 2. For any bipartite state $\varrho$ }
\begin{equation*}
K_{D}(\varrho) \leq E_{C}(\varrho).
\end{equation*}
This immediately implies $K_{D}(\varrho) \leq E_{F}(\varrho)$
since $E_{F}$ majorises $E_{C}$.
\section{Conclusions}
%

We have proven that any $d$--dimensional private state, i.e.,
state that contains $\log d$ bits of perfectly secure key is
distillable and we have provided the explicit LOCC operations of
distillation protocol. This result has already been applied by
other authors \cite{small} in a proof of unconditional security
with small distillable entanglement. We have also provided the
analogous result for a multipartite version of $d$--dimensional
private state which is a part of general scheme for distillation
of multipartite key \cite{MultiCryptoBE}. The results imply
immediately that although from bound entanglement one can produce
arbitrary secure bit of key (Eve's knowledge can be made
arbitrarily small), the latter can never be perfectly secure
(Eve's knowledge can not be made equal to zero). Finally
exploiting the structure of eigenvectors of p-bits we have
provided an elementary alternative proof of the fact that
entanglement cost is an upper bound on the amount of distillable
key of any quantum state.

\acknowledgments R. A. thanks Maciej Demianowicz for fruitful
discussions. This work was prepared under the (solicited) Polish
Ministry of Scientific Research and Information Technology grant
no. PBZ-MIN-008/P03/2003 and by EC grant RESQ, contract no.
IST-2001-37559.
%

%


\begin{thebibliography}{0}
\bibitem{Append}
Nonegativity of $a_{s}$ $(s=1,2)$ stems from the fact that
$\gamma^{(2)}$ is a state. To see that $\eta\le \sqrt{a_{1}a_{2}}$
one uses Cauchy-Schwarz inequality obtaining
%
\begin{eqnarray*}
\eta
&=&\left|\pf{1}U_{0}\sqrt{\rho_{A'B'}}\sqrt{\rho_{A'B'}}U_{1}^{\dagger}\pv{2}\right|\nonumber\\
&\le&\sqrt{\pf{1}U_{0}\rho_{A'B'}U_{0}^{\dagger}\pv{1}}\nonumber\\
&&\times\sqrt{\pf{2}U_{1}\rho_{A'B'}U_{1}^{\dagger}\pv{2}}\nonumber\\
&=&\sqrt{a_{1}a_{2}}.
\end{eqnarray*}
%
Since $\eta>0$, 
both numbers $a_{s}$
must be strictly positive.
\bibitem{LOCC}
By LOCC protocol one means any LOCC operation that is
trace-preserving.
\bibitem{vNEntropy}
For a given state $\Xi$ its von Neumann entropy reads $S_{\rm
vN}(\Xi)=-\Tr\Xi\log\Xi$.
\end{thebibliography}
\end{document}